\documentclass[19pt, preprint]{aastex}
\usepackage{graphicx}
\begin{document}

\title{Formation of Large Scale Coronal Loops Interconnecting Two Active Regions Through Gradual Magnetic Reconnection and Associated Heating Process}

\author{Guohui Du\altaffilmark{1}, Yao Chen\altaffilmark{1*}, Chunming Zhu\altaffilmark{2}, Chang Liu\altaffilmark{3}, Lili Ge\altaffilmark{4}, Bing Wang\altaffilmark{1}, Chuanyang Li\altaffilmark{1}, Haimin Wang\altaffilmark{3}}

\altaffiltext{1}{Shandong Provincial Key Laboratory of Optical
Astronomy and Solar-Terrestrial Environment, and Institute of
Space Sciences, Shandong University, Weihai 264209, China;
yaochen@sdu.edu.cn}
\altaffiltext{2}{Physics Department, Montana State University, Bozeman, MT 59717-3840, USA}
\altaffiltext{3}{Space Weather Research Laboratory, New Jersey Institute of Technology, University Heights, Newark, NJ 07102-1982, USA}
\altaffiltext{4}{Beijing Institute of Spacecraft Environment Engineering, Beijing 100094, China}

\noindent \textbf{Abstract}
Coronal loops interconnecting two active regions, called as interconnecting loops (ILs), are prominent large-scale structures in the solar atmosphere. They carry a significant amount of magnetic flux, therefore are considered to be an important element of the solar dynamo process. Earlier observations show that eruptions of ILs are an important source of CMEs. It is generally believed that ILs are formed through magnetic reconnection in the high corona ($>$150--$200"$), and several scenarios have been proposed to explain their brightening in soft X-rays (SXRs). Yet, the detailed IL formation process has not been fully explored and the associated energy release in the corona still remains unresolved. Here we report the complete formation process of a set of ILs connecting two nearby active regions, with successive observations by \emph{STEREO-A} on the far side of the Sun and \emph{SDO} and \emph{Hinode} on the Earth side. We conclude that ILs are formed by gradual reconnection high in the corona, in line with earlier postulations. In addition, we show evidence supporting that ILs become brightened in SXRs and EUVs through heating at or close to the reconnection site in the corona (i.e., through direct heating process of reconnection), a process that has been largely overlooked in earlier studies on ILs.

\noindent \textbf{1. Introduction}

Coronal loops interconnecting two active regions (ARs) are prominent large-scale structures frequently observed in the solar atmosphere. They are called as interconnecting loops (ILs) in general, or as trans-equatorial loops (TLs) if the two ARs are located on the opposite sides of the solar equator. These structures were first observed by \emph{Skylab} in soft X-rays (SXRs; \citealt[][]{Chase1976}). Since then, many studies have been conducted to investigate their observational characteristics and formation mechanisms.

According to the statistical study of \citet[][]{Pevtsov2000} with \emph{Yohkoh} data, about one third of ARs exhibit TLs. Therefore, ILs or TLs are not rare phenomena. They have attracted much attention mainly due to their important roles in the solar dynamo process and origin of solar eruptions. ILs (or TLs) with a significant poloidal component of magnetic field are regarded as evidence of the conversion of toroidal to poloidal component of the solar magnetic field, a critical step required in the Babcock-Leighton solar dynamo model (\citealt{Babcock1961, Leighton1969}; see also \citealt{Jiang2007}). Regarding their relation to the origin of solar eruptions, in particular, coronal mass ejections (CMEs), the statistical study by \citet[][]{Zhou2006} showed that nearly 40\% of all halo CMEs observed from March 1997 to December 2003 have sources closely related to TLs. \citet[][]{Wang2007} found that TLs associated with halo CMEs from NOAA AR 10696 present a positive acceleration, indicating a significant influence of TLs on CME dynamics.

Earlier studies suggested that these large-scale loops are formed by reconnection between field lines extending from ARs to the high corona. Pieces of evidence in support of this suggestion include cusp-like features above ILs, significant brightening in the SXR cusp region and along the whole loop structure (indicating presence of high-temperature and dense plasmas), and the dynamical and magnetic evolution of ARs and the nearby coronal holes (CHs) \citep[e.g.,][]{Svestka1977, Tsuneta1996, Bagal2000}.

During the life time of ILs, both transient and persistent brightenings in SXRs can be frequently observed. The chromospheric evaporation \citep[CE;][]{Neupert1968} was proposed to interpret SXR brightenings of ILs \citep[e.g.][]{Svestka1979}, in a way similar to that involved in the standard CSHKP picture of solar flares (\citealt{Carmichael1964, Sturrock1966, Hirayama1974, Kopp1976}). According to this scenario of IL brightenings, plasmas are mainly heated and brought into the ILs by the reconnection-induced CE process. Generally speaking, two different types of CE, explosive evaporation and the gentle evaporation, have been investigated, and are thought to be associated with different rates of energy injection through non-thermal energetic electrons and/or heat conduction (e.g., \citealt{Doschek1980, Feldman1980, Fisher1985a, Fisher1985b, Fisher1985c, Fisher1987, Mason1986, Brosius2004, Harra2005, Milligan2006, Tian2015}). In a case study on a TL that lights up in SXRs, \citet{Harra2003} reported blue shifts in the O V emission line around the TL footpoints, suggesting the presence of the CE process.

Brightenings of ILs were also interpreted as a result of magnetic field enhancement at their footpoints. The field enhancement may be due to flux emergence. \citet{Svestka1976} suggested that the increase of magnetic field strength at the footpoint leads to an enhanced amplitude of Alfv$\acute{e}$n waves. This might increase the wave dissipation rate and result in enhanced plasma heating and loop brightening. According to \citet{Svestka1976}, the dissipation is mainly due to the interaction of counter-propagating waves and thus concentrated at the loop top region. Although this suggestion seems to be consistent with the observations that ILs are often connected to newly emerging magnetic polarities, and that their brightenings often appear first at the top of ILs, especially during the early stage of ILs \citep[e.g.,][]{Svestka1979}, observational evidence indicating the presumed role of Alfv$\acute{e}$n waves has not been reported.

Note that these earlier studies mainly using SXR imaging data that suffer from discontinuous temporal coverage, poor spatial resolution, and low cadence. Studies using data from instruments on board \emph{Yohkoh} and the\emph{ Solar and Heliospheric Observatory} (\emph{SOHO}; \citealt{Domingo1995}) are in general consistent with the earlier suggestion that the reconnection high in the corona plays a critical role in the formation and brightening of ILs \citep[e.g.,][]{Bagal2000, Yokoyama2009, Yokoyama2010}. \citet{Yokoyama2009, Yokoyama2010} suggested that the reconnection between field lines from an AR and its nearby coronal hole is important to the formation of TLs. They presented a scenario of TL formation, in which some pre-existing seed magnetic fields, in the form of large-scale loops produced by reconnection between the AR and the coronal hole, are necessary. In their scenario, it is the eruption of these seed lines and subsequent flare-like reconnection that lead to the bright TLs. They assumed that the CE induced by the energy deposition from reconnection is the major process supplying high-temperature dense plasmas to the loops.

Despite the significant contributions of these earlier studies to the understanding of ILs, the formation process of ILs around the reconnection site high in the corona has not been observed directly. Limb events, such as that studied here, are helpful to reduce the interference from strong emissions of the underlying bright ARs. A combined analysis of multi-wavelength data covering a broad range of temperature is also necessary to reveal the thermodynamics of the high-corona ILs (and the relevant reconnection). In addition, it is reasonable to question whether such reconnection high in the corona, where the magnetic field strength is expected to be much weaker than that in the low corona, can still induce a prominent CE process to cause the in-general large-scale and long-term (in hours) IL brightening. The Atmosphere Imaging Assembly \citep[AIA;][]{Lemen2012} on board the \emph{Solar Dynamics Observatory} \citep[\emph{SDO};][]{Pesnell2012} has seven EUV passbands to image plasmas at different temperatures from 20,000 K to about 20 MK, with high spatial (0.6\H{ }pixel$^{-1}$) and temporal (12 s) resolutions. This provides a nice opportunity to reexamine the detailed high-corona reconnection process relevant to the formation and brightening of ILs.

\noindent \textbf{2. OBSERVATIONAL DATA AND EVENT OVERVIEW}\label{sec2}

The formation process of ILs of our event is observed by \emph{SDO}/AIA on the northeastern limb of the solar disk. The connection between ARs and the relevant magnetic configuration can be revealed about two days later with AIA and the Helioseismic and Magnetic Imager \citep[HMI;][]{Schou2012} on \emph{SDO}.

In Figure 1, we show images recorded by AIA at 171 {\AA} for the large-scale ILs (with one such loop indicated by a dotted curve in Figure 1a) at 23:39 UT on 2015 December 12 and by HMI for the underlying magnetic configuration of NOAA ARs 12469 and 12470. Both ARs are in the northern hemisphere. The AR 12469 is in its decay phase, which is characterized by weak and diffusive magnetic polarities, with a small and compact leading negative spot. The AR 12470 is still in its growing stage, undergoing a significant flux emergence; later it forms a major strong negative polarity followed by a diffusive positive polarity. The green dotted curve superposed onto the magnetogram delineates the overall structure of the ILs observed by AIA at almost the same time (see panel a). This shows that the ILs connect the following positive polarity of AR 12469 to the leading negative polarity of AR 12470. To verify this, the two footpoints of the ILs observed at the time of panel a are rotated to the time of the magnetogram shown in panel c (23:40 UT on 2015 December 14) at the nominal rate of solar rotation. The rotated footpoints are indicated with two green plus signs in panel c. They are consistent with the mentioned polarities of the sunspots.

The height of the ILs at the time of Figure 1a (23:39 UT on December 12) is estimated to be $\sim$150" above the solar disk, and the distance between the two footpoints is about 120". The ILs possess a clear poloidal component, as the angle between the line connecting the footpoints of the ILs and the longitudinal direction is about 60$^\circ$. This means that the ILs investigated here contain a significant poloidal polarity of the magnetic field. As introduced before, ILs with a poloidal component are an important element of the solar dynamo process.

The multi-temperature coverage of AIA enables us to analyze the heating-cooling process and loop dynamics involved in the IL formation. The ILs can be seen in most EUV passbands of AIA. Yet, their formation process is best observed in the 94 {\AA} passband (Fe XVIII; 6 MK), and the relevant dynamics of coronal loops and nearby radial structures are best observed in the 171 {\AA} passband (Fe IX; 0.6 MK). Therefore, we mainly analyze the AIA data in these two passbands. Data obtained with other passbands are also considered.

The AR 12469 appears in the \emph{SDO} field of view (FOV) at around 13 UT on December 11, and the AR 12470 appears $\sim$19 hours later ($\sim$8 UT on December 12). Both ARs emerge from below the photosphere on the backside of the Sun, before they rotate into the \emph{SDO} FOV. It is fortunate that their emerging process is recorded by the Extreme Ultraviolet Imager (EUVI; Wuelser et al. 2004) on board the \emph{Solar TErrestrial RElations Observatory A} (\emph{STEREO-A}; Kaiser et al. 2008) at all its four passbands. During the time of the event, \emph{STEREO-A} is about 167$^\circ$ behind the Earth. Data at 304 {\AA} (He II) and 195 {\AA} (Fe XII; 1.5 MK) of EUVI are analyzed here since they have higher temporal resolution (10 and 5 minutes, respectively) than the other two passbands (171 {\AA} and 284 {\AA}), and their combination can reveal the plasma properties at different temperatures ($\sim$10$^5$ to 10$^6$ K).

The ILs are also visible in SXR images from the X-Ray Telescope \citep[XRT;][]{Kosugi2007} on board the \emph{Hinode} spacecraft. This means that the plasmas within the ILs are hot and dense enough to emit in SXRs. This connects our study to those earlier ones that were mainly based on SXR data. We analyze the XRT SXR data of this event, which are available in the full-disk synoptic mode with a low cadence (twice per day, at about 06:00 UT and 18:00 UT; \citealt{2016SoPh..291..317T}) and a relatively high spatial resolution ($\sim$2" per pixel).

\noindent \textbf{3. DATA ANALAYSIS AND RESULTS}\label{sec3}

\noindent \textbf{3.1 EUVI/\emph{STEREO-A} observation of the emergence of the ARs}

In Figure 2, we present EUVI images at three times in 195 {\AA} and 304 {\AA} to show the emergence of the two ARs and relevant activities. An accompanying animation is available. It can be seen that there exists a small CH before the emergence of the AR 12469, which starts around 5 UT on December 05 from a location close to the southern border of the CH. With the AR emergence, significant activities including loop dynamics and brightenings are present. In about two days, the AR expands significantly. The southern border of the CH appears to retreat in response to the growth of the AR, suggesting that the CH and AR interact actively, likely through the interchange reconnection \citep{Crooker2002}. This may lead to the transport of open flux from the CH to the southeastern side of the AR, and thus partially account for the presence of the dark CH-like area there, as seen from panel c of this figure.

The AR 12470 starts to emerge at around 01 UT on December 7 from the southeastern part of the AR 12469. Similarly, its emergence is also accompanied by significant brightenings and loop dynamics (see white arrows in Figure 2). Transient large loops interconnecting the two ARs can be observed from 12--18 UT on December 09. After that, these large loops cannot be clearly recognized. The two ARs rotate to the solar limb on December 12 before entering into the FOV of \emph{SDO} (and \emph{Hinode}). The specific locations of the three spacecraft provide an almost complete coverage of the event.

\noindent \textbf{3.2 XRT/\emph{Hinode} and \emph{SDO} Observation of ILs and the associated reconnection process}

In Figure 3, we show four synoptic X-ray images obtained by XRT using the thin-Be filter. Before the time of panel a (05:49 UT on December 11), only the area above or within the active regions is bright in SXRs on the northeastern limb, while there is no signature of any high-lying loop structure. Later, there appears an arcade structure in the XRT data (see white arrows in panels b and c) above the ARs. As can be seen from these two panels (17:51 UT on December 11 and 06:04 UT on December 12), the top of the arcade is brighter than its lower internal part. The arcade top is located at about 100" above the limb, and the distance of the two footpoints is estimated to be $\sim$150". The arcade in panel d appears to be higher than those observed earlier. The appearance time, location, and morphology of the X-ray arcade are consistent with those observed by AIA in various EUV passbands for the ILs (see below). This indicates that the X-ray arcade and the EUV ILs originate from the same structure.

The formation process of ILs is well observed by \emph{SDO}/AIA. In Figure 4, we show AIA images observed at 94 {\AA}, 171 {\AA}, and 304 {\AA} at four different times. An accompanying animation is also available. The three left panels (05:49 UT) are for the onset of IL formation. At this time, low-lying bright emission is present above the AR 12470 that is still behind the solar disk. In the high corona, there appears some diffuse emission at 94 {\AA} (see the white arrow in panel a). On the northern side, a set of prominent radial structures is observed at 171 {\AA} (see the black arrow in panel e). They could be open or large-scale closed field lines. Some bright structures (see the green arrow in panel i) are observed above the ARs at 304 {\AA}.

The formation process of ILs can be viewed from the time sequence of images presented in this figure and the accompanying animation. The process lasts for $\sim$1.5 days, from $\sim$05 UT on December 11 to $\sim$23 UT on December 12. Starting from 05 UT on December 11, loop structures continuously rise into the corona from the northeastern limb of the solar disk. It is likely that, these loops emerge from the ARs, especially the southern AR 12470, which is in the early developing stage. The rising loops interact with the bright radial structures (see the 171 {\AA} images). The interaction results in the formation of a new group of bright loops which become the ILs. During this IL formation process, the radial structures vanish gradually, and a clear cusp-like structure with a straight stalk can be observed (see panel b). After the time of panel b (17:51 UT on December 11), the radial structures adjacent to the newly-formed loops become more and more curved, forming the cusp structure prominent at 171 {\AA}. The upper part of the cusp is much dimmer at 171 {\AA} compared to other parts of the cusp. This dim region at 171 {\AA} corresponds to the bright cusp-shaped emission at 94 {\AA}, as seen from top panels. This indicates that the upper part of the cusp (where the 94 {\AA} emission is enhanced) contains hot plasmas with temperatures as high as 4--6 MK.

To further show the relative location of enhanced emissions at 171 {\AA} and 94 {\AA} (also 193 {\AA} and 211 {\AA}), in Figure 5 we present time-distance plots along the slice S1. The white line in each panel of this figure delineates the bright front at the corresponding wavelength of the present panel, and the dashed line represents the front given in the preceding panel. It is seen that the bright front (i.e., the location with emission intensities that are significantly higher than that from the undisturbed area at similar altitudes) measured in hotter passbands is always higher than that measured in cooler passbands. The front at 94 {\AA} is much higher than those observed in the other three wavelengths (by $\sim$30--$60"$), while the front at 211 (193) {\AA} is only slightly higher (by $\sim$10--15") than the one at 193 (171) {\AA}.

As seen from the AIA data, the plasmas emitting at 94 {\AA} in the high corona originate from the upper part of the cusp. There appears a long thin structure (see the red arrow in Figure 4c) originating from the cusp. To better visualize the details of loop dynamics, we use the Multi-scale Gaussian Normalization \citep[MGN;][]{Morgan2014} method to further process the data at 94 {\AA} and 171 {\AA}. The processed images are shown in Figure 6 and the accompanying animation.

It can be seen that the heated plasmas observed at 94 {\AA} are born persistently at the cusp region in the high corona, and then they move downwards. To show this, in Figure 7a we plot the time-distance image along the slice S3. Despite the overall slowly-rising trend of the region with enhanced emission at 94 {\AA} (see Figure 5d), there appear lots of sunward-flowing loop structures, as delineated with the dashed lines in Figure 7a. Their speeds are estimated to be $\sim$  3 km~s$^{-1}$. These observations of AIA at 94 {\AA} provide direct evidence of plasma heating around the cusp region. Figure 7b is the time-distance plot along the slice S2. Loops from the southern AR keep rising through the process with an upward velocity $\sim$  8--20 km~s$^{-1}$ as estimated from the time-distance plot. It is interesting to note that the tips of some 171 {\AA} structures show small-amplitude turbulent oscillations. They may be driven by the same process that heats the plasmas and produces the enhanced emission at 94 {\AA}.

In Figure 8, we show light curves at various AIA passbands within the small box defined in Figure 4h. The intensity peaks are present first at 94 {\AA} (14:00 UT), and then at 335 ($\sim$17:00 UT), 221 (18:40 UT), 193 (18:50 UT), 171 (19:08 UT), and 131 (19:17 UT) {\AA}, in the sequence of declining temperature according to the AIA response functions. In Figure 5, we have presented the locations of bright fronts observed at 94, 211, 193, and 171 {\AA}. Putting the two observations together, the AIA data indicate an on-going cooling process, starting from the highest region that appears first at 94 {\AA}. Note that the flux profile at 335 {\AA} is very broad in comparison to others, and as a result, its maximum is not well defined. This is because that this passband has a broad response in temperature.

The cooling process eventually results in an emission enhancement at 304 {\AA}, as seen from the movie accompanying Figure 4. The 304 {\AA} emission first appears at the top of the large loops, and forms an arc-shaped structure. To reveal the relevant flow pattern, in Figure 9 we presents distance-time plots during two intervals along the arc-shaped structures observed at 304 {\AA}. It can be seen that, the flows originate from the top of the arc with a clear downward motion at both sides. This means that the enhanced emission at 304 {\AA} is a result of the cooling process as described above.

According to the movie accompanying Figure 4, after December 12 the radial structures on the northern side vanish entirely. Almost at the same time, loop activities, including formation of new loops and brightening at loop tops, appear to be weaker. The overall brightness of the whole loop system decreases correspondingly, indicating a substantial reduction of high-corona activities. Later, the CH is not recognizable any more from the AIA data.

In summary, the above observations from \emph{SDO}/AIA reveal the following aspects of IL formation. Firstly, there presents an overall cusp-like structure in the high corona ($\gtrsim$ 150--200" above the disk), where plasmas are heated to a temperature of several MK (emitting at 94 {\AA}). The heated plasmas then cool down to temperatures corresponding to passbands of 335, 221, 193, 171, and 131 {\AA}, forming the bright IL loops observed at 171 {\AA}. Plasmas eventually observed at 304 {\AA} flow downward from the top of the arcade, in a way similar to coronal rains that are sometimes observed after flares. This indicates significant plasma cooling and condensation. The source of condensation can be provided by the pre-reconnection loops that are successively rising from the ARs and those pre-existing radial structures. In addition, downward motions of loop-like structures can be observed at 94 {\AA}. This is similar to downward contracting arcades (an evidence of outflow of reconnection in the high corona) observed after solar flares \citep[see, e.g.,][]{Liu2013, Wu2016}. Comparing the large loop structure observed at 94 {\AA} by AIA (Figure 4c) and those in SXR by XRT (Figure 3d), we see that the enhanced X-ray emission in the high corona corresponds to the region with bright 94 {\AA} emission. This indicates that the two types of emission enhancement have the same physical origin.

From the above observations, we suggest that the reconnection in the high corona and the resultant heating process around the reconnection site lead to the formation of ILs and their brightening in both SXRs and EUVs.

\noindent \textbf{3.3 A SCENARIO FOR THE FORMATION OF ILs}

The three spacecraft (\emph{STEREO}, \emph{Hinode}, and \emph{SDO}) provide a multi-wavelength and continuous coverage of the event, including emergence of the two ARs, interaction between the AR 12469 and the nearby CH, and formation and thermodynamics of the ILs. In this section, we present a combined analysis of these data using four schematics shown in Figure 10.

The schematics represent different stages of the event. In panel a, we show the AR 12469 that just emerges through the photosphere (see Figure 2a), together with the field lines extending outward radially from the northern coronal hole. In panel b, the other AR 12470 that is emerging from the southeastern side is also drawn. The interchange reconnection takes place between the AR 12469 and the CH. This results in a partial transport of long radial magnetic field lines towards the southern part of the AR 12469, as well as retreat of the CH boundary and growth of the AR (see Figure 2b). In panel c, reconnection in the high corona occurs between radial field lines and loops rising from the AR 12470. This forms an overall cusp-like structure. Loops (indicated in red) appear underneath the cusp, as a result of downward contraction of post-reconnection field lines and cooling of plasmas there. Panel d shows further reconnection between the radial field lines and the loops that continuously rise from the younger AR, as well as the growth of newly-born loops that interconnect the two ARs. As mentioned, the radial field lines that reconnect are likely rooted at the eastern border of the AR 12469. They may consist of field lines that have been transported from the CH through the above-mentioned interchange reconnection, and large-scale closed field lines rooted within the AR 12469. During the process, the area of the CH decreases and its border retreats as a result of its reconnection with the AR 12469. After the event, the CH fades away in the FOV.

\noindent \textbf{4. Conclusions and Discussion}\label{sec4}

This study presents observations of the formation of ILs and relevant brightenings in SXRs and EUVs. Multi-wavelength data from three spacecraft (\emph{STEREO}, \emph{SDO}, and \emph{Hinode}) are analyzed. It is found that ILs are formed and heated at or near the reconnection site in the high corona. This is mainly supported by the AIA data obtained at 94 {\AA}, which show that high-temperature plasmas appear at the cusp-like region and then move sunward and cool down to lower temperature at lower altitudes. The reconnection takes place between field lines of loops rising from the ARs and some open-like radial magnetic structures, and it lasts for more than one day. This indicates a highly-asymmetric and gradual reconnection process. The nearby CH plays a role in forming these radial magnetic structures, particularly, during its earlier interaction (on the far side of the Sun) with the emerging AR. Since no non-thermal emission and rapid motion of plasmas (and loop structures) are observed during the process, we suggest that the reconnection forming the ILs in the high corona as investigated here takes place in a very gradual manner, and mainly converts magnetic energy into thermal energy.

In earlier studies, observations of IL formation and brightenings are mainly performed using the \emph{Skylab} data \citep[e.g.,][]{Chase1976, Svestka1977-1, Svestka1979}, which are discontinuous in temporal coverage and have poor spatial resolution and low cadence. Later studies analyzed the \emph{Yohkoh} data in SXRs \citep[e.g.,][]{Tsuneta1996, Bagal2000, Pevtsov2000, Chen2006, Yokoyama2009, Yokoyama2010} and \emph{SOHO} data in EUVs \citep[e.g.,][]{Zhou2006, Wang2007}. Compared to the imaging instruments (\emph{SDO}/AIA) used in this study, those data suffer from an insufficient coverage of plasma temperatures, which prevents previous authors from observing the thermodynamic details at the high corona reconnection site. In particular, the heating and cooling processes relevant to the formation and evolution of ILs have not been well observed.

Nevertheless, it has been suggested by, e.g., \citet{Tsuneta1996} and \citet{Yokoyama2009, Yokoyama2010}, that the CE process, act in a way similar to what proposed in the standard picture of solar flares, may be important for the heating and brightening of ILs (or TLs). Since the reconnection leading to IL formation occurs in the high corona ($>$150--200" above the disk) where the magnetic field strength is much weaker in comparison to that in the low corona, it may not be energetic enough to drive the CE process that can lead to the large-scale and long-term (relative to the time scale of usual solar flares) IL brightening. Mainly using data at 94 {\AA} that represents a major high-temperature channel of AIA, we have shown evidence supporting that the heating and brightening of ILs at SXRs and EUVs are a result of direct heating at or near the reconnection site.

In addition, \citet{Yokoyama2009, Yokoyama2010} concluded that ILs are formed from the eruptions of the so-called seed magnetic field lines, which themselves are large-scale loops connecting ARs and nearby CHs. In our event, no signatures of eruptions of large-scale loops are found, instead, the reconnection in the high corona leading to the formation of ILs is driven by the persistent rising and expanding motion of loops mainly attributed by the younger AR. The role of the CH in our event is to provide open-like radial field lines through earlier interchange reconnection with the older AR. These radial structures are quasi-steady. The difference of our event from those reported by Yokoyama \& Masuda may indicate that ILs can be formed through different processes, either related or not related to large-scale loop eruptions, yet the high-corona reconnection seems to be always necessary.

The \emph{STEREO}/SECCHI data are produced by an international consortium of the NRL, LMSAL and NASA GSFC (USA), RAL and University of Birmingham (UK), MPS (Germany), CSL (Belgium), IOTA, and IAS (France). \emph{SDO} is a mission of NASA's Living With a Star Program. \emph{Hinode} is a Japanese mission developed and launched by ISAS/JAXA, with NAOJ as a domestic partner and NASA and STFC (UK) as international partners, and is operated by these agencies in cooperation with ESA and NSC (Norway). We thank Dr. Jiang Jie for helpful discussion on solar dynamo models. This work was supported by NNSFC grants 41331068, 11790303 (11790300), and NSBRSF 2012CB825601.

\begin{figure}
\epsscale{1.}
\includegraphics[width=1.0\textwidth]{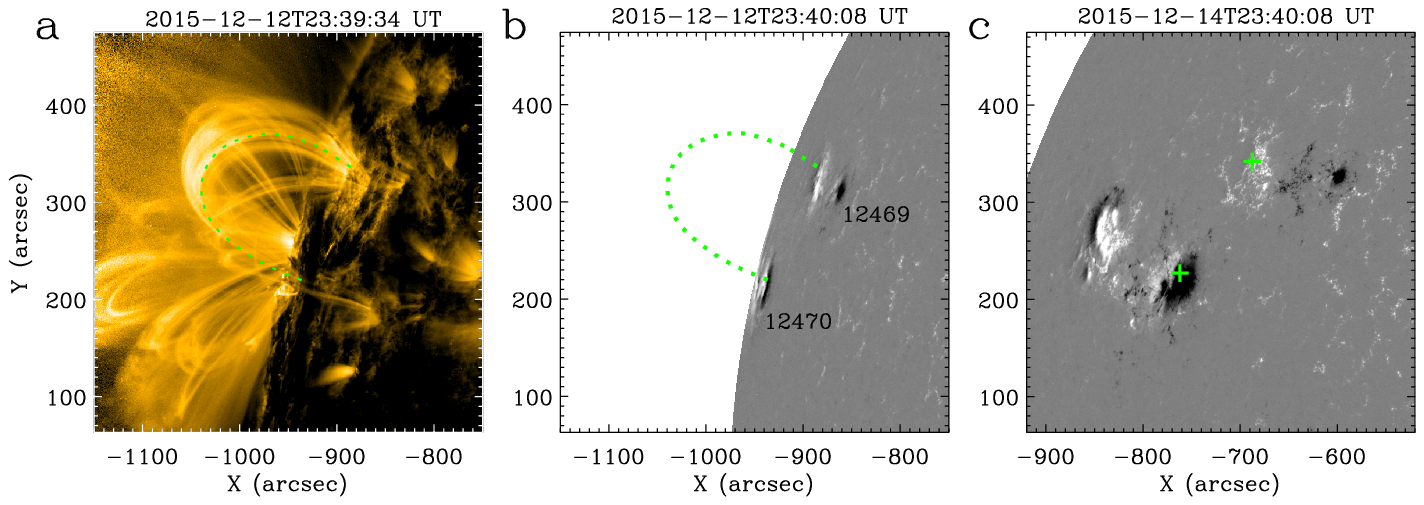}
\caption{
HMI/\emph{SDO} magnetograms for the magnetic configuration of the ARs
(NOAA 12469 and 12470, panels b and c). AIA/\emph{SDO} image at 171 {\AA} for
the large-scale ILs (panel a). The green curves superposed onto panel a and b are given by delineating the overall morphology of the ILs at almost the same time. The plus signs in panel (c) denote the footpoints of the ILs, after been rotated from the time of panel a to panel c at the nominal rate of rotation of the Sun.
}\label{Fig1}
\end{figure}

\begin{figure}
\epsscale{1.}
\includegraphics[width=1.0\textwidth]{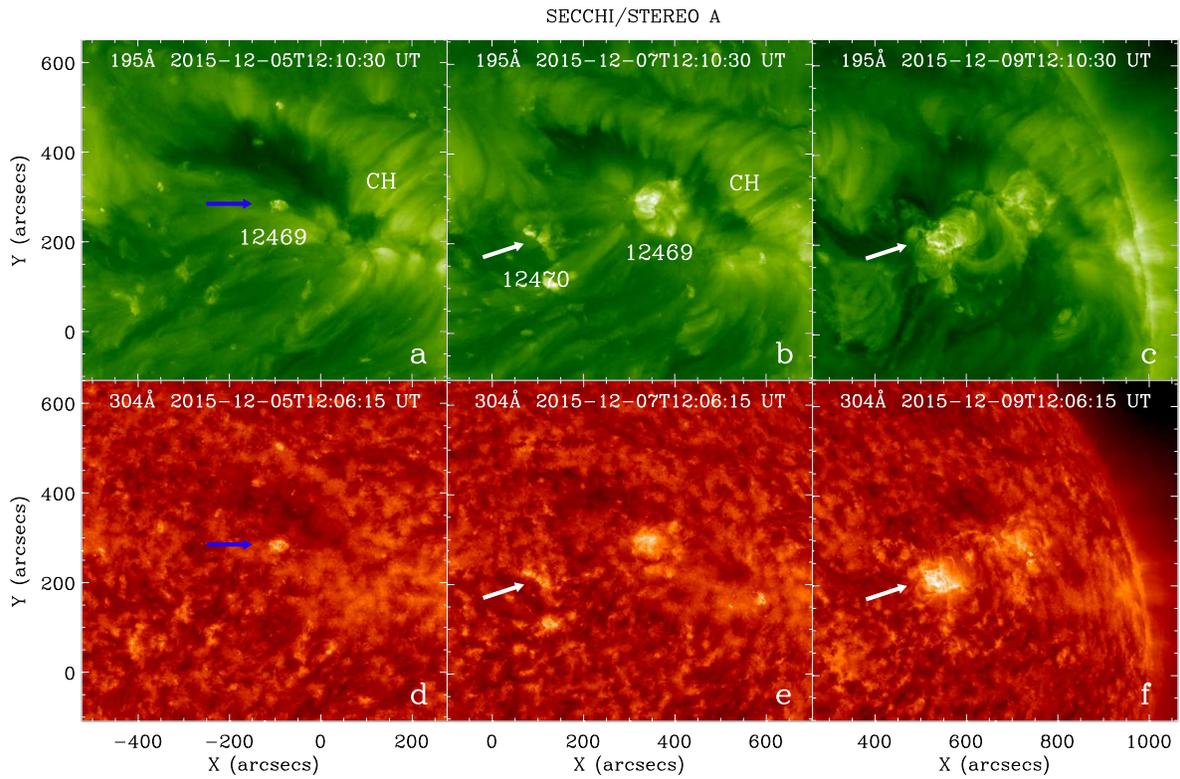}
\caption{
EUVI/\emph{STEREO} A images at 195 {\AA} and 304 {\AA} to show the emergence
of the two ARs, the CH and relevant activities. The blue arrows in panel a and d point to the emerging
AR 12469. The white arrows in panel b and e point to the emerging
AR 12470.
}\label{Fig2}
\end{figure}

\begin{figure}
\epsscale{1.}
\includegraphics[width=1.0\textwidth]{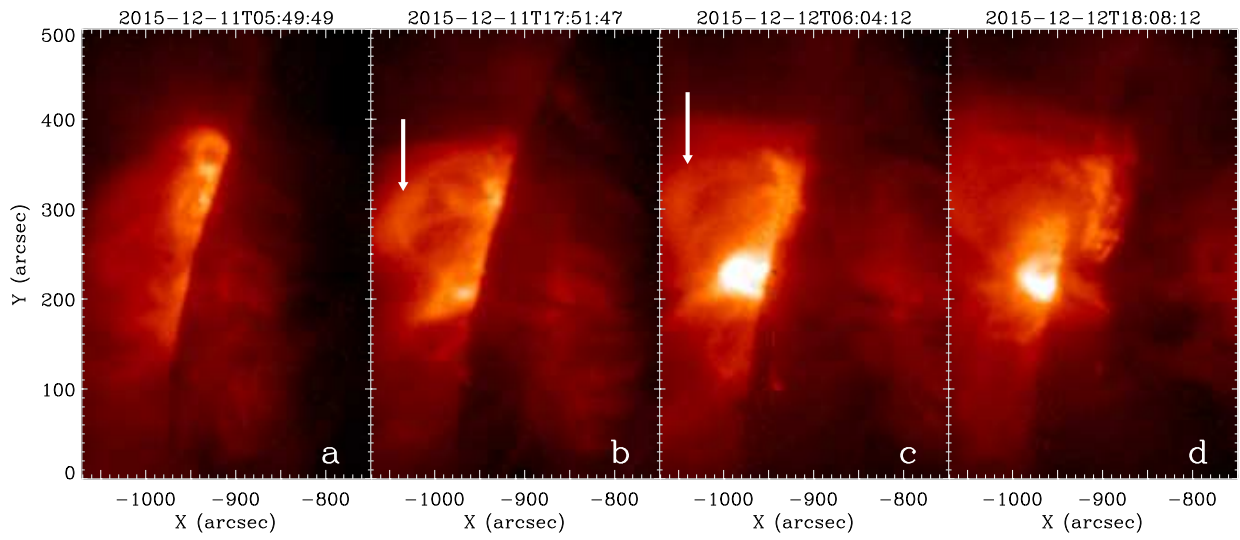}
\caption{
SXR images of the formation of ILs recorded by XRT/\emph{Hinode}, in the full-disk synoptic mode. See text for details.
}\label{Fig3}
\end{figure}

\begin{figure}
\epsscale{1.}
\includegraphics[width=1.0\textwidth]{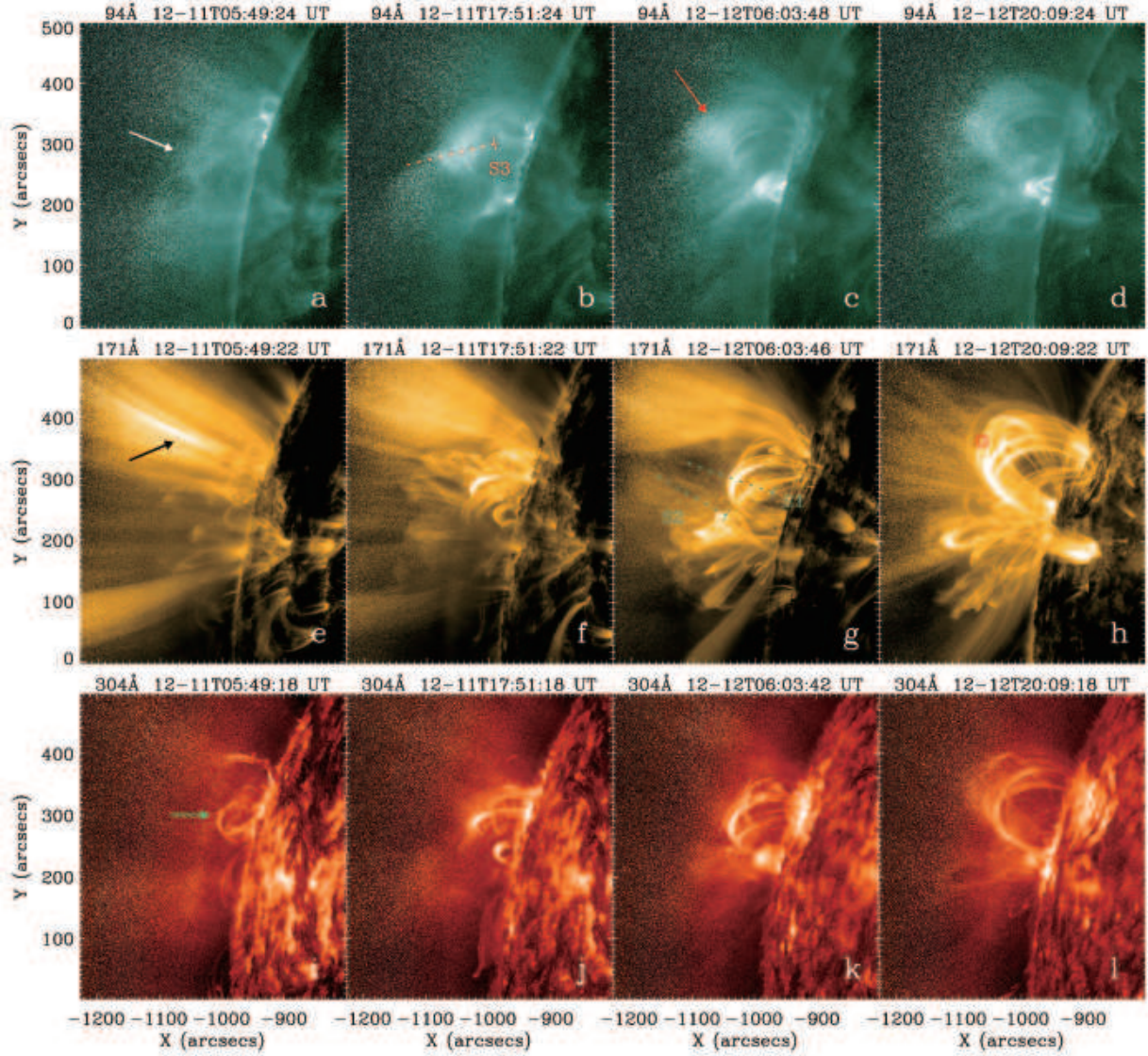}
\caption{
AIA/\emph{SDO} images at 94, 171, 304 {\AA} passbands for the formation process of ILs.
The left 3 panels are for the onset of ILs formation, the middle 6 panels are for the formation
process of ILs, and the right 3 panels are images observed at the end of the ILs formation process.
Dashed lines denote the positions of slices S1, S2, and S3, along which the distance map are generated and shown in Figure 5 (S1), and Figure 7 (S2 and S3). The red box shown in panel h is used to deduce the light curves at various AIA passbands (see Figure 8).
(An animation of this figure is available.)
}\label{Fig4}
\end{figure}

\begin{figure}
\epsscale{1.}
\includegraphics[width=0.5\textwidth]{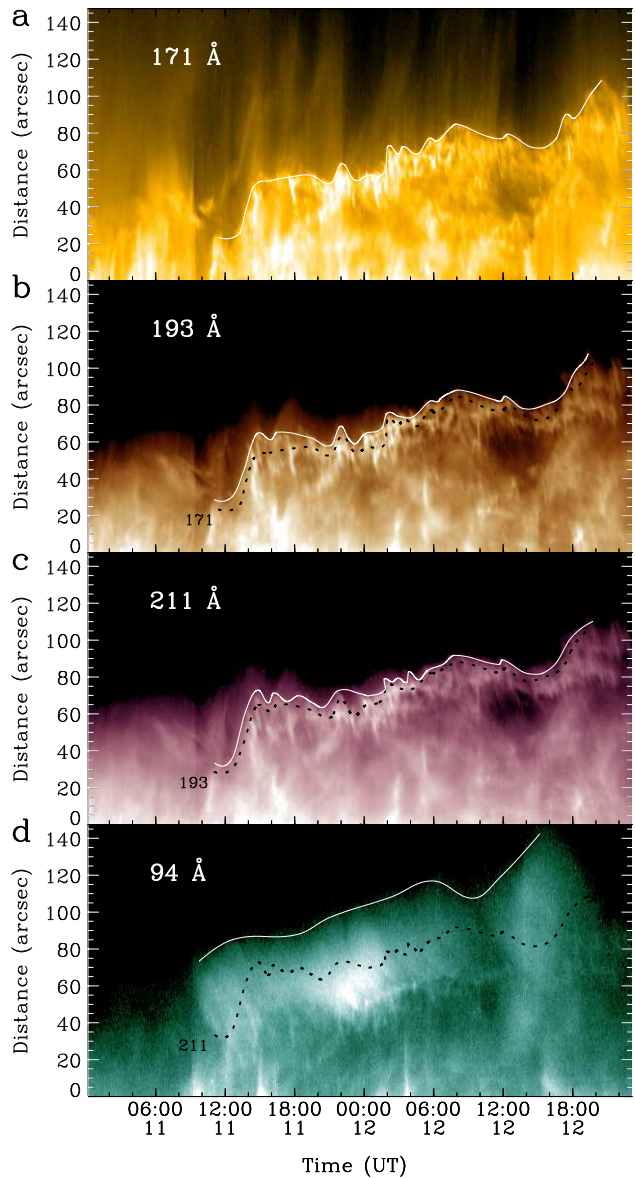}
\caption{
Distance-time plots at 171, 193, 211, and 94 {\AA} along S1 (see Figure 4g). These panels are ordered in the sequence of increasing formation
temperatures according to the AIA response functions. In each panel, the white solid curve delineates the bright front
at the present wavelength, the black dashed line represents the bright front delineated in the preceding panel.
}\label{Fig5}
\end{figure}

\begin{figure}
\epsscale{1.}
\includegraphics[width=1.0\textwidth]{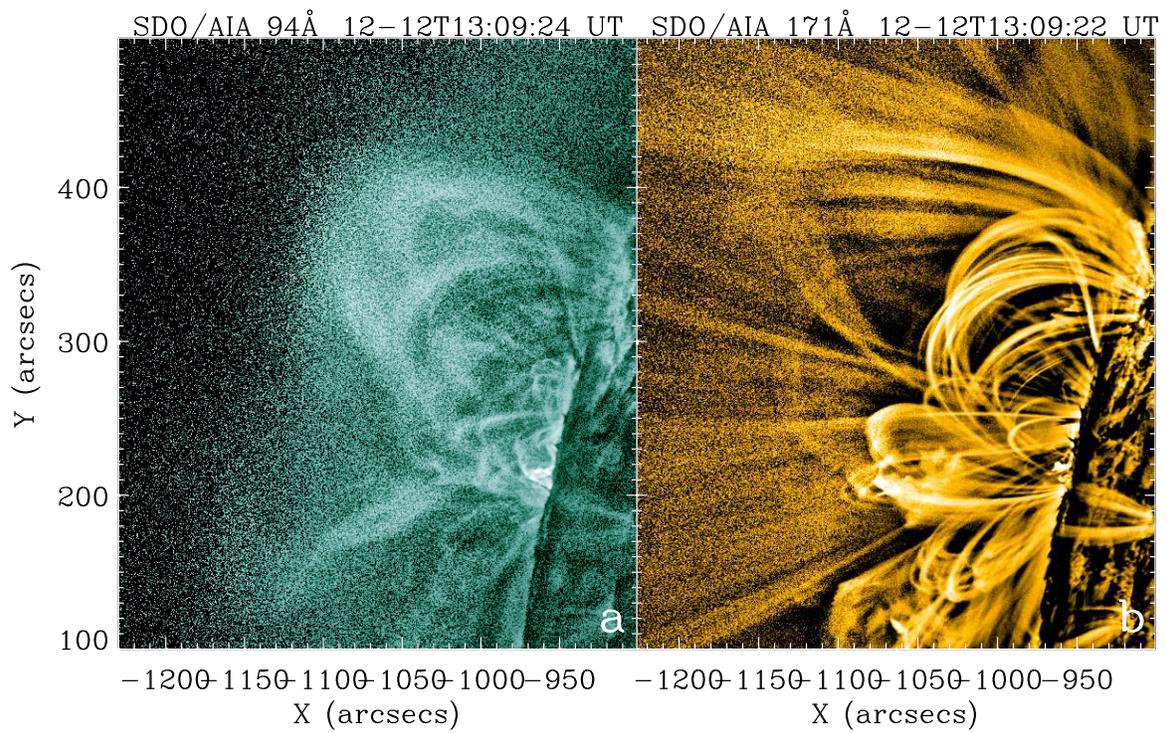}
\caption{
AIA/\emph{SDO} images at 94 and 171 {\AA} that are further processed by the Multi-scale Gaussian
Normalization method to better visualize details of the loop dynamics, and the heating process.
(An animation of this figure is available.)
}\label{Fig6}
\end{figure}

\begin{figure}
\epsscale{1.}
\includegraphics[width=1.0\textwidth]{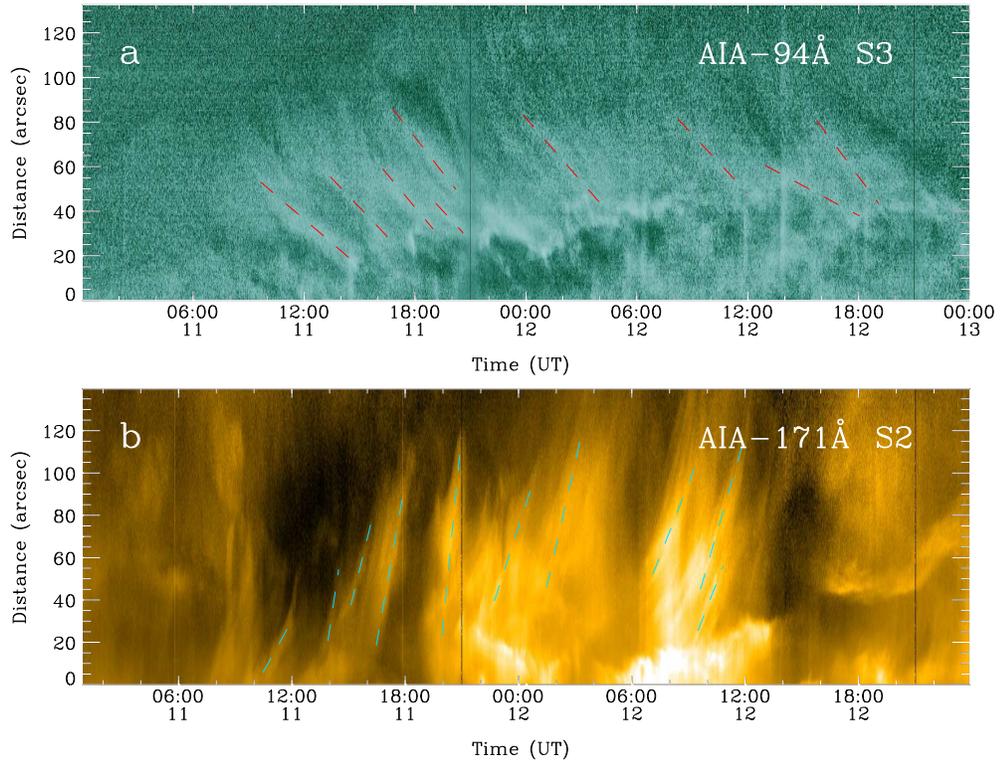}
\caption{
Distance-time plots at 171 {\AA} and 94 {\AA} along the slice S2 and S3 that have been shown in Figure 4. The dashed lines are tracers of the rising cool loop structures in the upper panel (171 {\AA}), and the downward-flowing hot structures in the lower panel (94 {\AA}). These lines are used to measure the speeds of relevant motion.
}\label{Fig7}
\end{figure}

\begin{figure}
\epsscale{1.}
\includegraphics[width=1.0\textwidth]{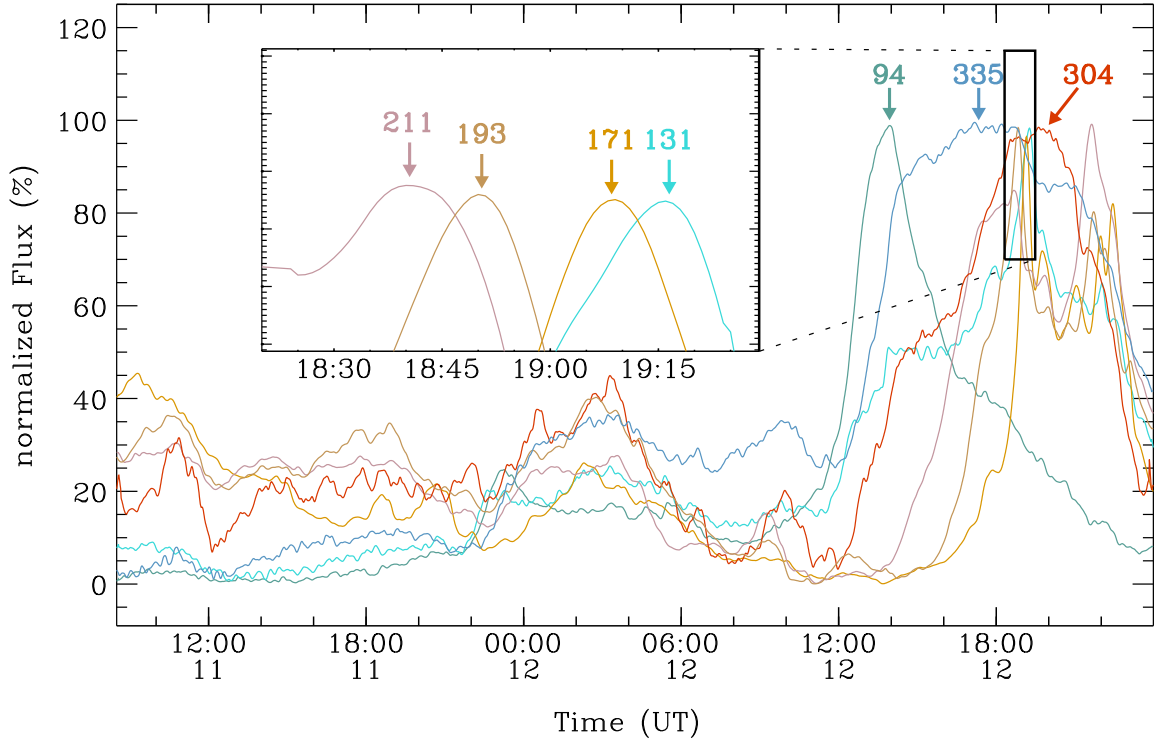}
\caption{
Light curves at various AIA passbands within the small box defined in Figure 4h. The inset shows an enlarged version of the small squared region from 18:20 - 19:30 UT. The maximum of
intensity is present first at 94 {\AA} at 14:00 UT, and then at 335 (~17:00 UT), 221 (18:40 UT), 193
(18:50 UT), 171 (19:08 UT), and 131 (19:17 UT){\AA}, in sequence of declining temperature, according
to the AIA response functions.
}\label{Fig8}
\end{figure}

\begin{figure}
\epsscale{1.}
\includegraphics[width=0.5\textwidth]{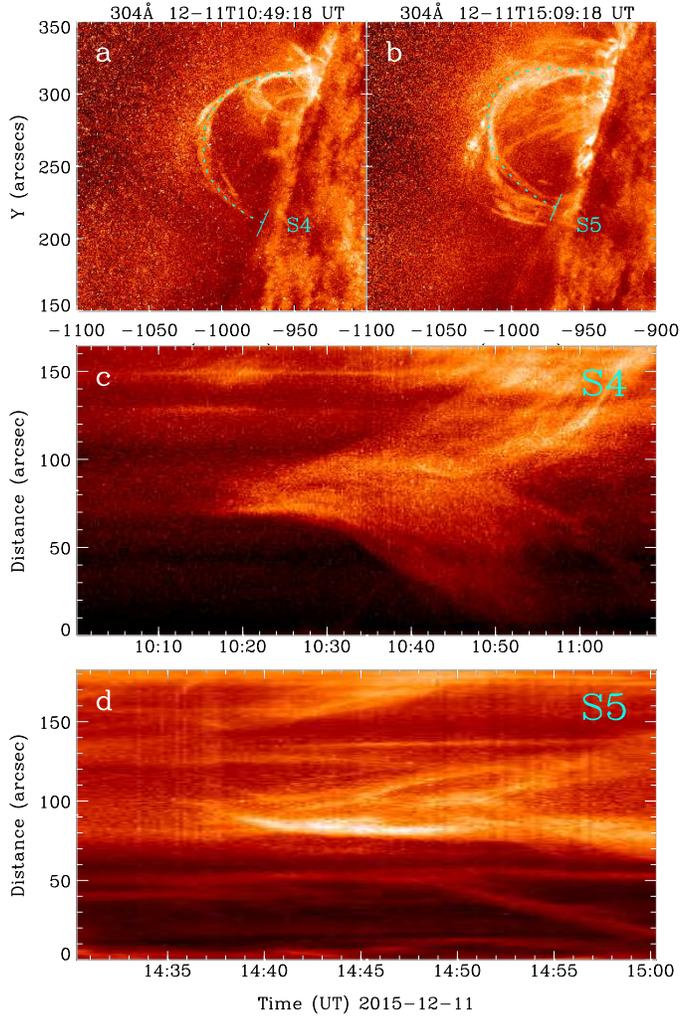}
\caption{
Distance-time plots along two curved slices (S4 and S5) that are defined in upper panels of  AIA/\emph{SDO} images at 304 {\AA}. A clear
downward motion at both sides can be observed from the lower two panels (c and d).
}\label{Fig9}
\end{figure}

\begin{figure}
\epsscale{1.}
\includegraphics[width=1.0\textwidth]{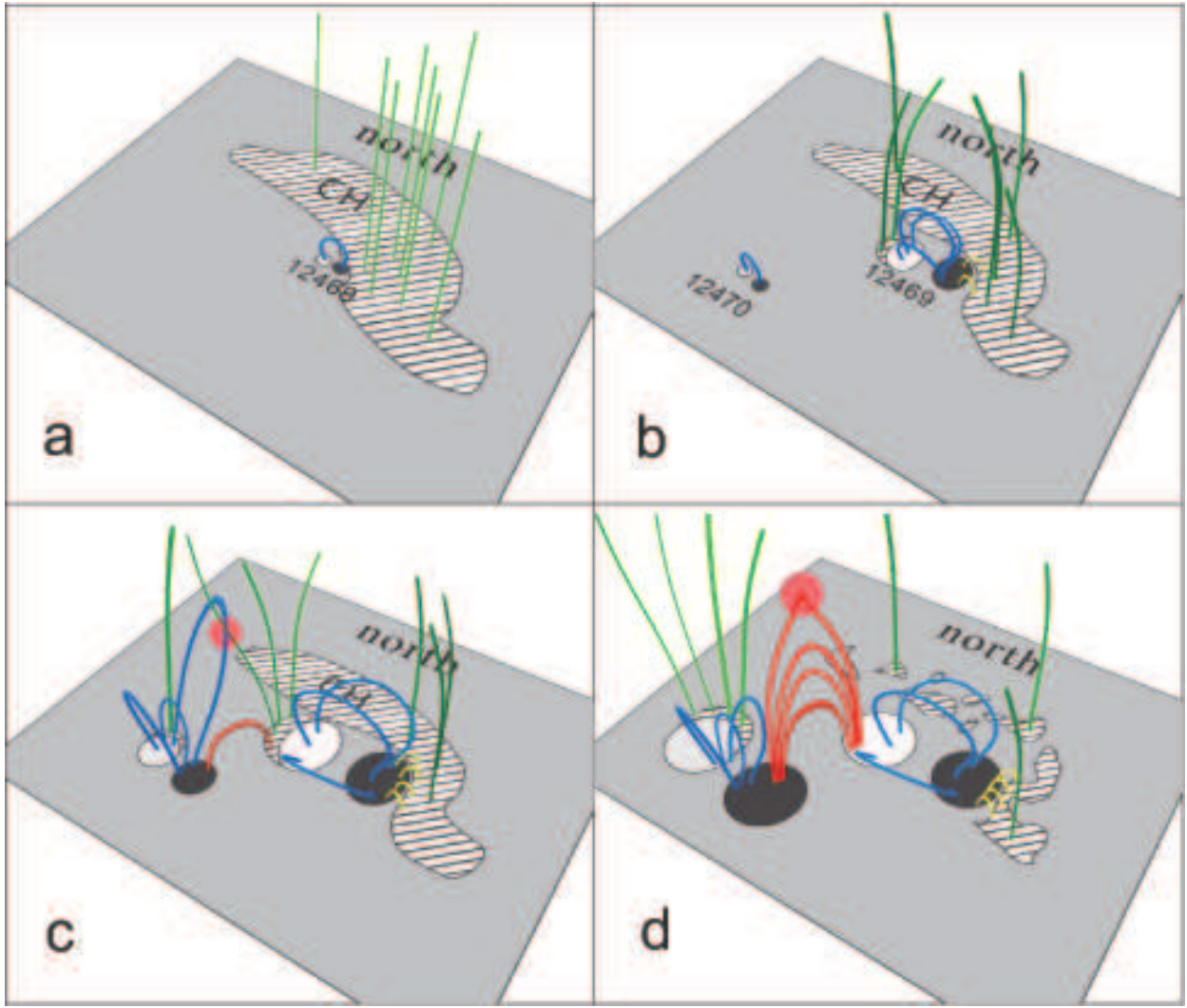}
\caption{
Schematics to show different stages of the IL formation process. Green and blue lines represent the field lines from the CH and the two ARs. The low-lying yellow loops represent the loops generated by the interchange reconnection between the CH and the nearby AR, the high lying red loops are for the newly formed large scale ILs interconnecting the two ARs (i.e., the ILs). Red solid-circular regions represent the reconnection sites.
}\label{Fig10}
\end{figure}

\bibliography{ref}
\end{document}